\documentclass[prl,twocolumn,showpacs,superscriptaddress]{revtex4}
\usepackage{graphicx}% Include figure files
\usepackage{dcolumn}% Align table columns on decimal point
\usepackage{bm}% bold math
\usepackage{mathtools}
\usepackage{color}
\usepackage{general_latex}

\begin{document}

\title{Remnant quantum resources of collapsed macroscopic quantum superpositions}

\author{T.J.~Volkoff}
\email{adidasty@gmail.com}
\affiliation{17682 W. Shields Ave., Kerman, CA 93630}

\date{\today}

\begin{abstract}
In this Letter, we consider the collapse of a macroscopic quantum superposition occurring due to the measurement which optimally distinguishes its branches. Given a macroscopic  superposition of $N$ spin-1/2 particles, we use such a Helstrom measurement to construct the local unitary operator which maximizes the usefulness of the superposition for Heisenberg-limited phase estimation (i.e., with quantum Cram\'{e}r Rao bound proportional to $1/N$). In contrast, the collapsed state is not useful as a probe for phase estimation below the standard quantum limit. For the case $N=2$, we compute the entanglement entropy of the collapsed state and show that it is reduced below that of the initial superposition when the superposition is macroscopic. We consider the remnant quantum resources of collapsed hierarchical Schr\"{o}dinger cat states to show that collapsed macroscopic superpositions can still be useful for ultraprecise quantum metrology.
\end{abstract}

\pacs{03.67.Bg,03.65.Ta}

\maketitle

In quantum mechanics, an optimal measurement of ``which path'' or trajectory information of a quantum object is traditionally associated with probabilistic collapse (according to the Born rule) of the quantum state of the particle to a state vector corresponding to a definite path \cite{tarasov,schloss}. If the paths are completely distinguishable, i.e., if they are represented by a set of orthogonal vectors in Hilbert space, all quantum coherence between the paths is lost during the collapse. A paradigmatic example is provided by the collapse of the entangled state of kinematic momentum and spin angular momentum of a spinful particle to a definite classical trajectory upon measurement carried out by a Stern-Gerlach apparatus \cite{plastino}.  However, the projection of a superposition onto one of its branches does not necessarily result in a state devoid of quantum mechanical utility. As a counterexample, measurement of ``which path'' information of engineered entangled states has been used to create macroscopic quantum superpositions, e.g., photonic Schr\"{o}dinger cat states, in cavity QED devices \cite{schoelkopfcat,harochedavidovich}.

Many well-known photonic Schr\"{o}dinger cat states exhibit small, but nonzero overlap between the two branches; examples include the even and odd coherent states \cite{dodonovart} and entangled coherent states \cite{manko}. These example photonic superpositions lie in a large class of quantum superposition states of the form \cite{volkoff4} \begin{equation}\ket{\psi}= {(\mathbb{I} + U^{\otimes N})\ket{\phi}^{\otimes N} \over \sqrt{2+2\text{Re}z^{N}}} \label{eqn:general} \end{equation} (with $\ket{\phi}$ in the single-mode Hilbert space $\mathcal{H}$, $U$ unitary, $\mathbb{I}$ the identity operator on $\mathcal{H}^{\otimes N}$, and $z:= \langle \phi \vert U \vert \phi \rangle$) that are labeled ``macroscopic superpositions'' according to measures based on: 1) the optimal distinguishability of their branches under measurements of small subsystems \cite{whaleyjan},  and 2) their usefulness as probes for ultraprecise phase estimation (i.e., having quantum Cram\'{e}r-Rao bound scaling inversely to the total number of photons) of an appropriate unitary evolution \cite{dur}.  In this Letter, we motivate and define a ``collapsing measurement'' (called CM from here on) for equally-weighted, two-branch (orthogonal or nonorthogonal) quantum superpositions and show that it coincides with the optimal measurement for distinguishing between the branches in a pure state binary distinguishability setting. For the case of orthogonal branches, the CM reduces to the traditional notion described above.  We then analyze the remnant entanglement entropy and metrological usefulness of the states resulting from an application of a CM to a macroscopic quantum superposition having the form of Eq.(\ref{eqn:general}) with the aim of answering dual questions: 1) What are the quantum resources required to perform a CM on a macroscopic quantum superposition? and 1') If a CM has been carried out on the superposition, what quantum resources remain?

For an equal-weight superposition of two nonorthogonal single-mode states $\ket{\phi}$ and $U\ket{\phi}$, where $U$ is a unitary operator, a measurement which collapses the superposition should produce completely distinguishable outcomes with equal probability. In addition, taking for inspiration the classical ``alive'' and ``dead'' states existing after a measurement that collapses Schr\"{o}dinger's famous cat superposition \cite{schrodinger}, it is preferable that the CM should produce as outcomes states that are very close to $\ket{\phi}$ and $U\ket{\phi}$ with equal probability.  We use example of the even coherent state, which takes the form of Eq.(\ref{eqn:general}), having $\ket{\phi}=\ket{\alpha}$ an oscillator coherent state, $U=e^{i\pi a^{\dagger} a}$, and $N=1$, to discuss two different quantum measurements, each satisfying one of the aspects of our notion of collapse:  

\begin{itemize}
\item Projective measurement onto states of definite energy, e.g., of the Hamiltonian $H=\hbar \omega a^{\dagger}a$. Such a measurement collapses $\ket{\psi_{+}}$, in the sense of producing completely distinguishable outcomes, however the outcomes do not occur with equal frequencies. Any possible outcome $\ket{n}$ exhibits the minimal value of $-1$ for Mandel's $Q$ parameter \cite{mandel} and thus exhibits non-Poissonian photon statistics, in contrast to the branches $\ket{\pm \alpha}$. 
\item Measurement containing $\ket{\pm \alpha}\bra{\pm \alpha}$, e.g., a homodyne measurement. Such a measurement produces the branch states as alternatives with equal probability, but the superposition has not been collapsed because the resulting states are not completely distinguishable, i.e., the measurement is not orthogonal.
\end{itemize}

We now define a single mode ($N=1$) CM which satisfies both aspects of our notion of collapse.  \underline{Definition}: A single mode CM for the superposition of Eq.(\ref{eqn:general}) is a POVM containing orthogonal projection operators $E_{+}$, $E_{-}$, and $E_{3}:= \mathbb{I} - E_{+}-E_{-}$ such that $\text{tr}(E_{\pm}\ket{\psi}\bra{\psi}) = 1/2$. 

Note that if the Hilbert space is two dimensional, the POVM contains only two elements, $E_{\pm}$. The outcomes of a CM are, by definition, completely distinguishable. But are the outcomes very close to $\ket{\phi}$ or $U\ket{\phi}$, e.g., for the case of the even coherent state, does the CM very nearly produce one of the classical states $\ket{\alpha}$ or $\ket{-\alpha}$? We answer affirmatively by finding the CM for $\ket{\psi}$ and proving that the outcomes are states which very close to either $\ket{\phi}$ or $\ket{U\phi}$ as long as $z \ll 1$. We take the overlap $\langle \phi \vert U \vert \phi \rangle = z$ to be real for simplicity. In the Hilbert space spanned by complex linear combinations of $\ket{\phi}$, $U\ket{\phi}$, the CM is uniquely defined by the requirements of: 1) $E_{\pm}^{2}=E_{\pm}$ (projectivity), 2) $\text{tr}(E_{\pm}\ket{\psi}\bra{\psi})=1/2$ (probabilistic symmetry) \footnote{If a quantum measurement is projective, then it is orthogonal. See Ref.\cite{holevo}.}. Solving the algebraic equations produces $E_{\pm} := \ket{\xi_{\pm}}\bra{\xi_{\pm}}$ with:
\begin{equation}
\ket{\xi_{\pm}} ={\sqrt{1-z} \mp \sqrt{1+z} \over 2\sqrt{1-z^{2}}}\ket{\phi} + {\sqrt{1-z} \pm \sqrt{1+z} \over 2\sqrt{1-z^{2}}}U\ket{\phi}  \label{eqn:xi}
\end{equation}
where we have taken $z$ to be a real number.
Clearly, for $z \ll 1$, $\ket{\xi_{-}}$ ($\ket{\xi_{+}}$) has nearly unit overlap with $\ket{\phi}$ ($U\ket{\phi}$).  It is useful to explicitly note that $\vert \langle \phi \vert \xi_{+} \rangle\vert^{2} = 1/2(1-\sqrt{1-z^{2}})$, $\vert \langle \phi \vert U^{\dagger}\vert \xi_{+} \rangle\vert^{2} = 1/2(1+\sqrt{1-z^{2}})$.

It is intriguing that the spectral decomposition of the self-adjoint operator  $\ket{\phi}\bra{\phi} - U\ket{\phi}\bra{\phi}U^{\dagger}$ takes the form $\sum_{i=\pm}\lambda_{i}E_{i}$, with $\lambda_{+} = -\lambda_{-}$ being real numbers, i.e., this operator defines a measurement observable for the CM. In addition, it is well-known that the largest probability of successfully distinguishing two quantum states $\rho$ and $\sigma$, present with equal \textit{a priori} probabilities in  a quantum channel, is obtained by applying a measurement (the Helstrom measurement) containing the spectral projections of the operator $\rho - \sigma$ to the channel \cite{helstrombook,fuchs}. Hence, we see that our present notion of CM coincides with the Helstrom measurement for the states $\ket{\phi}$ and $U\ket{\phi}$. The fact that the Helstrom measurement elements in Eq.(\ref{eqn:xi}) can be produced from $\ket{\phi}$ and $U\ket{\phi}$, respectively, by a unitary transformation appears in Refs.\cite{sasaki1,sasaki2} in the context of optimal quantum state discrimination.

\begin{figure}
\includegraphics[scale=0.45]{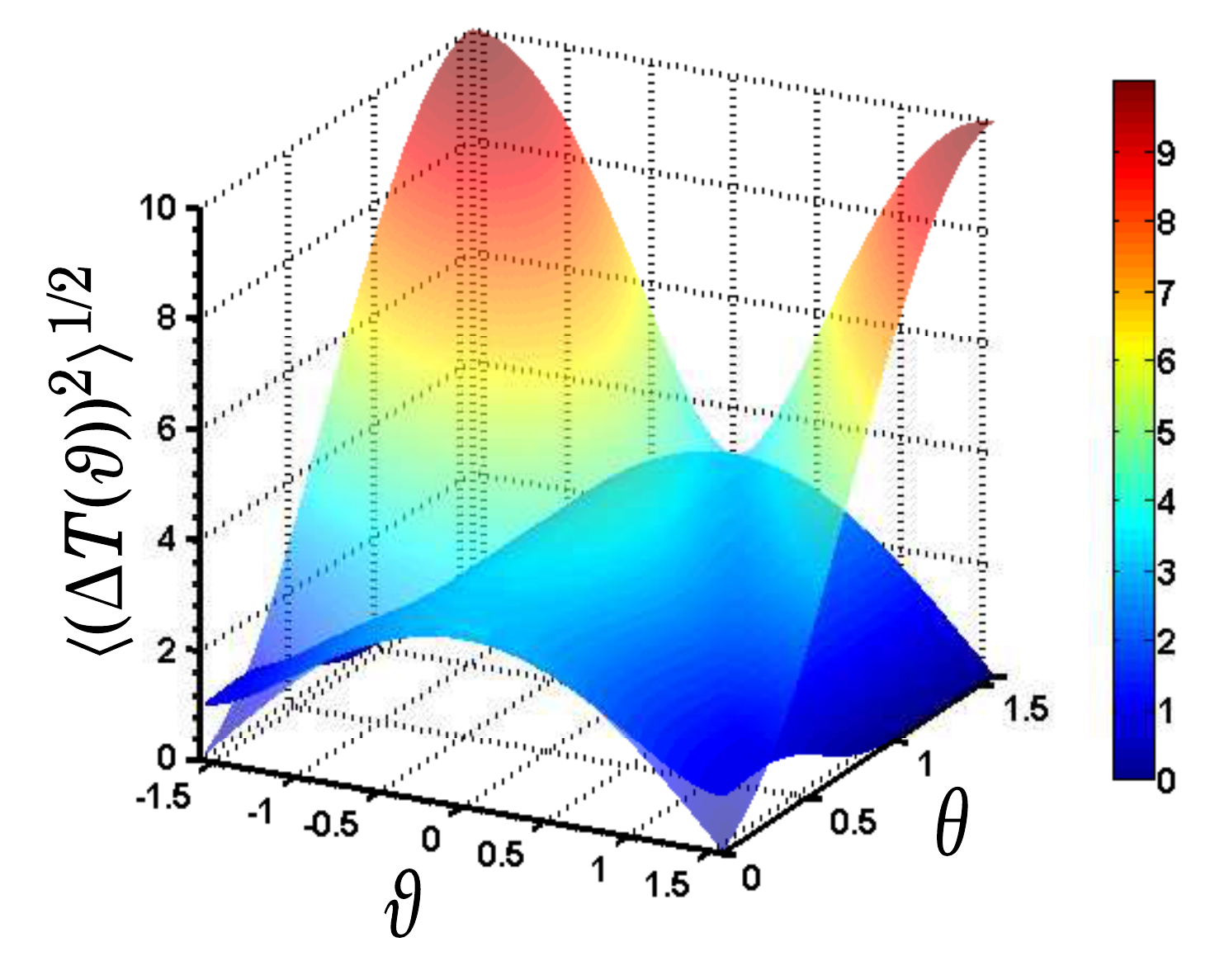}
\caption{The deviations of the 1-local spin operator $T(\vartheta)$ in the superposition $\ket{\psi(\theta)}$ and collapsed superposition $\ket{\Omega_{+}(\theta)}$, $\theta \in (0,\pi /2]$ for $N=10$ plotted from their analytical expressions. Note the asymmetry across the $\vartheta = 0$ axis, which is due to the direction of the relative rotation of the branches. Also, for small $\theta$, the collapsed superposition allows greater maximal deviation for $T(\vartheta)$ (over all $\vartheta$) than the initial superposition.\label{fig:var}}
\end{figure}

For the multimode ($N>1$) case, we define the CM as the Helstrom measurement distinguishing between $\ket{\phi}^{\otimes N}$ and $U^{\otimes N}\ket{\phi}^{\otimes N}$. This measurement is not separable \cite{duan}, and in the remainder of the paper, we explore the quantum mechanical resources of this measurement. One may object that a separable measurement with elements given by $E_{i_{1}}\otimes \ldots \otimes E_{i_{N}}$, $i_{k} \in \lbrace +,-,3 \rbrace$ produces equiprobable and completely distinguishable outcomes and so is consistent with the present notion of collapse. However, for large $N$, most of the possible outcomes have negligible overlap with both branches $\ket{\phi}^{\otimes N}$ and $U^{\otimes N}\ket{\phi}^{\otimes N}$, so the spirit of a three-outcome measurement with distinguishable output and large overlap with one of the branches is lost \footnote{This can be quantified by showing that the expected fidelity between a measurement outcome and either of the branches is $2^{-N}$.}. Hence, the possible outcomes of a CM applied to Eq.(\ref{eqn:general}) for $N>1$ are given by: \begin{eqnarray}
\ket{\Omega_{\pm}} &:=& {\sqrt{1-z^{N}} \mp \sqrt{1+z^{N}} \over 2\sqrt{1-z^{2N}}}\ket{\phi}^{\otimes N} \nonumber \\ &+& {\sqrt{1-z^{N}} \pm \sqrt{1+z^{N}} \over 2\sqrt{1-z^{2N}}}U^{\otimes N}\ket{\phi}^{\otimes N} .\label{eqn:collapse}
\end{eqnarray} That the optimal measurement for pure state binary distinguishability can be alternatively characterized by the two simple assumptions of a CM for a two-branch superposition is a unique feature of the two-branch case. It is natural to try to extend the definition of CM to equal weight superpositions of $m>2$ linearly independent pure states and to subsequently determine if the extended definition agrees with the optimal measurement for pure state $m$-ary detection \cite{helstrombook}. For an equally-weighted superposition of linearly independent states $\ket{\psi}\propto \sum_{j=1}^{m}\ket{\phi_{j}}$, a general definition of CM must provide for $m^2$ values $\langle \xi_{j}\vert \phi_{k}\rangle$. Extending the previous conditions: 1) $\vert \langle \xi_{j}\vert \psi\rangle\vert^{2}=1/m$, and 2) $\langle \xi_{j}\vert \xi_{k}\rangle = \delta_{jk}$ together provide $(m^{2}+m)/2$ constraints. The remaining $m(m-1)/2$ constraints can be physically motivated by the symmetry $\ket{\phi_{j}}\mapsto \ket{\phi_{s(j)}}$ of $\ket{\psi}$, with $s$ a permutation of $m$ letters. As we have seen, each pure state $\ket{\phi_{k}}\bra{\phi_{k}}$ is associated to a CM element $\ket{\xi_{k}}\bra{\xi_{k}}$ with which it has the largest (Hilbert-Schmidt) inner product. We demand that for all permutations $s$, the measurement $\lbrace \ket{\xi_{sk}}\bra{\xi_{sk}} \rbrace$ be the CM for $\ket{\psi_{s}}\propto \sum_{k}\ket{\phi_{sk}}$ such that $\langle \phi_{k} \vert \xi_{j} \rangle = \langle \phi_{sk} \vert \xi_{sj} \rangle$ for all $k$, $j$.  Proof that the three above conditions imply that the CM is the optimal $m$-ary pure state receiver \cite{kennedy,holevo2} with an appropriate cost function follows from basic considerations and is omitted.

For more complicated superpositions, e.g., involving continua of states, the connection between the notion of superposition collapse and optimal pure state distinguishability is less clear.

With the explicit form of a CM in hand, we compare two quantum resources, metrological usefulness and entanglement, of an initial macroscopic superposition to those of its collapsed image. We first consider the following $N$-mode spin-1/2 superposition having the form of Eq.(\ref{eqn:general}) with $\ket{\phi}=\ket{0}$, $U = e^{-i\theta \sigma_{y}}$:
\begin{equation}
\ket{{\Psi(\theta)}} = {\ket{0}^{\otimes N} + (\cos \theta \ket{0} + \sin \theta \ket{1})^{\otimes N} \over \sqrt{2+2\cos^{N}(\theta)}}.\label{eqn:spinsup}
\end{equation}
Such a superposition is considered macroscopic when $\theta \approx \pi/2$ \cite{cirac,dur}. According to the quantum Cram\'{e}r-Rao theorem \cite{holevo}, the maximal  precision (over all $\vartheta \in [0,\pi/2]$) obtainable by a pure quantum state of an $N$-site spin-1/2 chain used as a probe for phase estimation of the evolution generated by the 1-local Zeeman Hamiltonian $T(\vartheta) = {1\over 2}\sum_{i=1}^{N} \vec{\sigma}^{(i)}\cdot (\cos \vartheta,0,\sin \vartheta)$ is inversely proportional to $\text{max}_{\vartheta} \left( \langle (T(\vartheta) - \langle T(\vartheta) \rangle)^{2}\rangle \right)^{1/2}$. Note that the operator norm of $T(\vartheta)$ is $N$ for all $\vartheta$. This maximal deviation can be calculated by the method of, e.g., Ref.\cite{smerzi}. The deviation of $T(\vartheta)$ in $\ket{\psi(\theta)}$ and one of the outcomes, $\ket{\Omega_{+}(\theta)}$, of the CM is plotted in Fig. \ref{fig:var}. When the branches of $\ket{\psi(\theta)}$ are nearly orthogonal, the maximal precision over $\vartheta$ is inversely proportional to $N$, i.e., $\ket{\psi(\theta)}$ can serve as a probe for Heisenberg-limited metrology. On the contrary, the maximal precision when $\ket{\xi_{+}(\theta)}$ is used as probe is always inversely proportional to $\sqrt{N}$, i.e., the collapsed state never beats the standard quantum limit. For the state $\ket{\psi(\theta)}$, the maximal precision for phase estimation of evolution generated by $T(\vartheta)$ occurs for $\vartheta = -\theta$. A glance at the Hamiltonian $T(-\theta)$ confirms that it can be rewritten in the form: \begin{equation}
T(-\theta)=\sum_{i=1}^{N}(\ket{\xi_{-}(\theta)}\bra{\xi_{-}(\theta)} - \ket{\xi_{+}(\theta)}\bra{\xi_{+}(\theta)})^{(i)} 
\end{equation}
where $\ket{\xi_{\pm}(\theta)}\bra{\xi_{\pm}(\theta)}$ are the CM operators for the superposition in Eq.(\ref{eqn:spinsup}) with $N=1$. We arrive at the remarkable conclusion that a superposition having the form of Eq.(\ref{eqn:spinsup}) is most useful for phase estimation of unitary evolution generated by the Hamiltonian $\sum_{i=1}^{N}(E_{-}-E_{+})^{(i)}$, where $E_{\pm}$ are elements of the CM for the superposition $\propto \ket{0} + e^{-i\theta \sigma_{y}}\ket{1}$. The geometrical statement corresponding to this physical fact is that the Fubini-Study line element on the one-parameter path $\text{exp}(-it\sum_{i=1}^{N}(E_{-}-E_{+})^{(i)})\ket{\psi(\theta)}$ in projective Hilbert space \cite{bengt} is greater than on any other such path generated by a 1-local Hamiltonian having operator norm $N$. This has the important consequence that the time required for $\ket{\psi(\theta)}$ to evolve to any state on the path $e^{-iHt}\ket{\psi(\theta)}$, with $H=H^{\dagger}$ 1-local and having norm $N$, is minimal when $H=\sum_{i=1}^{N}(E_{-}-E_{+})^{(i)}$ \cite{volkoff4}.
\begin{figure}
\includegraphics[scale=0.35]{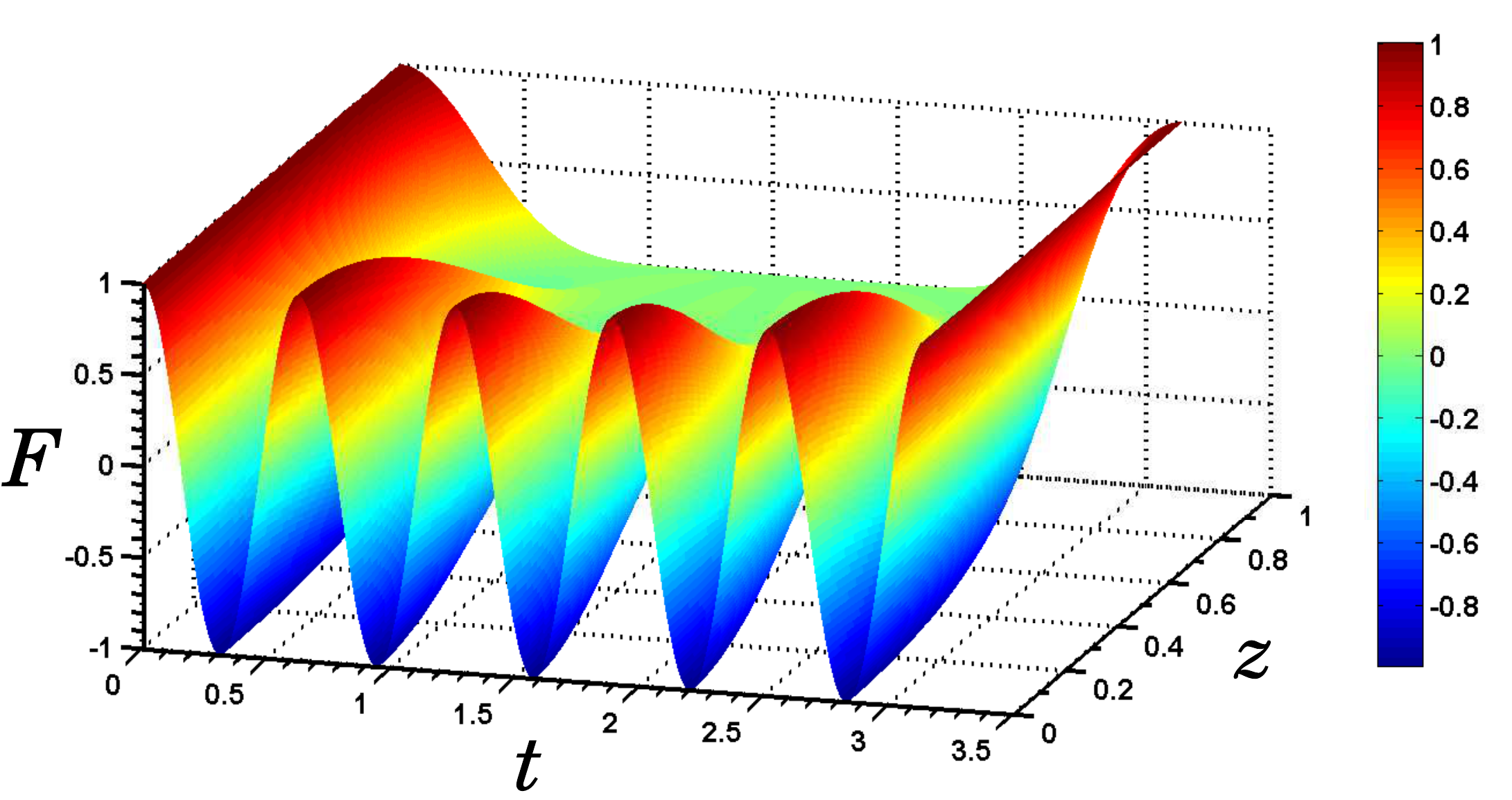}
\caption{Overlap $F:= \langle \psi \vert  \text{exp}(-i\omega t\sum_{i=1}^{N}(E_{-}-E_{+})^{(i)}) \vert \psi \rangle$ for $N=10$ and $\omega = 1$ plotted from the analytical expression with respect to $z=\langle \phi \vert U \vert \phi \rangle \in [0,1]$ and time $t\in [0,\pi]$.  \label{fig:evol}}
\end{figure}

 We demonstrate this maximal orthogonalization speed for the general state in Eq.(\ref{eqn:general}). The unitary time-evolution generated by $\omega \sum_{i=1}^{N} (E_{-}-E_{+})^{(i)}$ where $\omega$ is an arbitrary energy scale is given by:
\begin{eqnarray}
e^{-i\omega t \sum_{i=1}^{N} (E_{-}-E_{+})^{(i)}}&=& e^{-i\omega t \sum_{i=1}^{N} E_{-}^{(i)} }e^{i\omega t \sum_{i=1}^{N} E_{+}^{(i)} } \nonumber \\ &=& (e^{-i\omega t E_{-}})^{\otimes N}(e^{-i\omega t E_{+}})^{\otimes N} \nonumber \\ &=& (e^{-i\omega t} E_{-} + e^{i\omega t} E_{+})^{\otimes N}
\end{eqnarray}
where we have used the projection property $E_{\pm}^{2}=E_{\pm}$ of the CM, and have assumed a two dimensional Hilbert space for each mode. The overlap $F:= \langle \psi \vert  \text{exp}(-i\omega t\sum_{i=1}^{N}(E_{-}-E_{+})^{(i)}) \vert \psi \rangle$ is shown in Fig.\ref{fig:evol}. Note that $F$ only regains its initial value of 1 when $z=0$, i.e., the branches of $\ket{\psi}$ are orthogonal. The greater speed of orthogonalization occurring for $z=0$ as compared to $z=1$ is clear from the figure, reflecting the greater Fubini-Study line element for the macroscopic superposition. The experimental implementation of optimal distinguishing measurements of nonorthogonal spin-1/2 quantum states \cite{waldherr} indicates the feasibility of realizing the CM and the concomitant fast orthogonalization for macroscopic quantum superpositions of the form Eq.(\ref{eqn:spinsup}).

\begin{figure}
\includegraphics[scale=0.35]{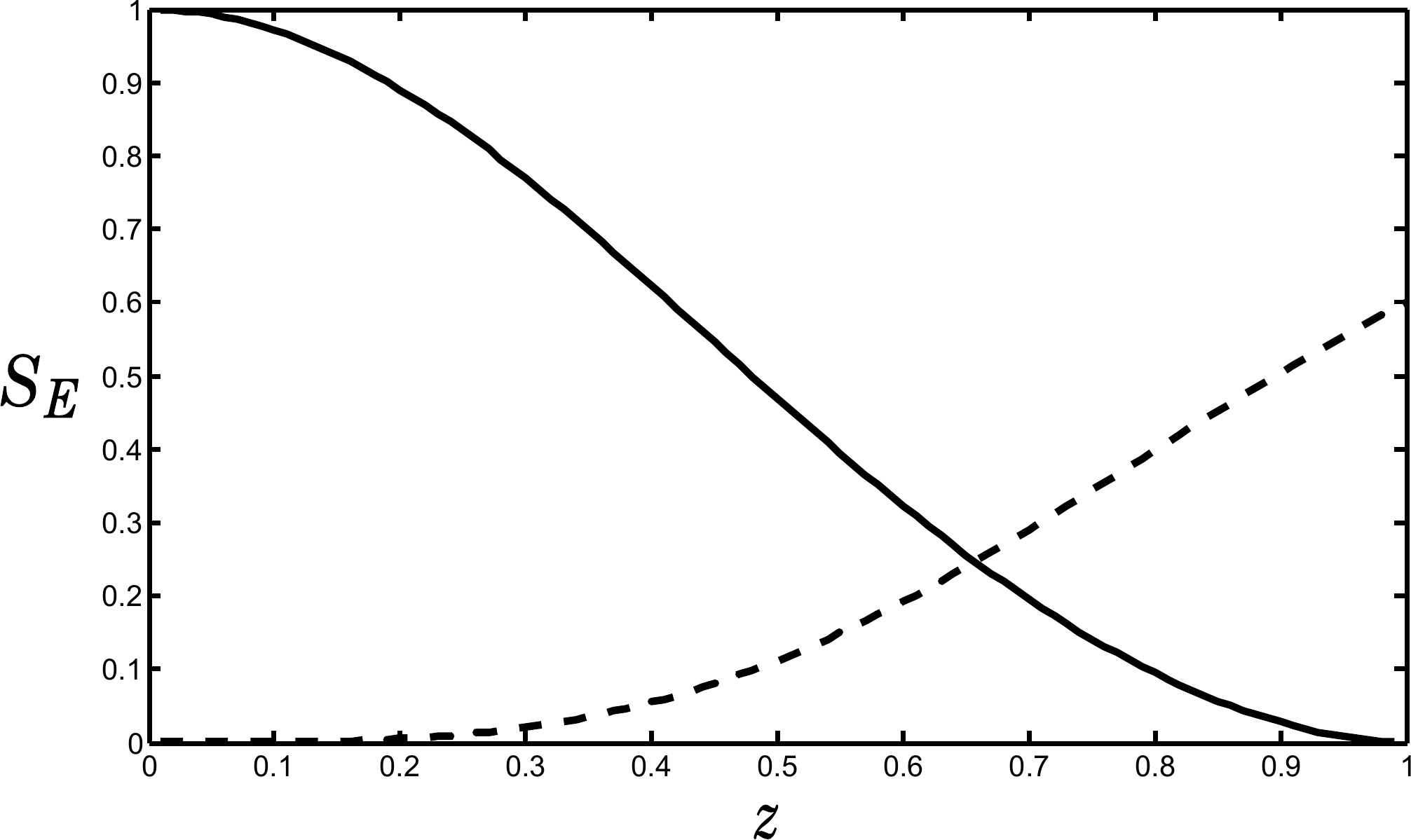}
\caption{Entanglement entropy of $\ket{\psi}$ (solid curve) and $\ket{\Omega_{+}}$ (dashed curve) for $N=2$ and $z \in [0,1)$. We have taken Boltzmann's constant as $k_{\text{B}}=1$. \label{fig:entang}}
\end{figure}

It is known that there exist entangled spin states that are not useful for metrology at precisions greater than the standard quantum limit \cite{smerzi}. Conversely, it has been shown that certain nonentangled states can beat the standard quantum limit for phase estimation of evolutions generated by certain nonlinear interactions \cite{beyondheisenprodstate}. Hence, it is important to compare the quantum resource of entanglement of macroscopic superpositions and the states produced probabilistically by a CM.  We consider the entanglement entropies \cite{weinberg} of the general state of Eq.(\ref{eqn:general}) for $N=2$ and a corresponding collapsed state of Eq.(\ref{eqn:collapse}). As can be seen in Fig. \ref{fig:entang}, the entanglement entropy of $\ket{\Omega_{+}}$ is greater than that of $\ket{\psi}$ for $z\gtrsim 0.6573$. The most important feature of Fig. \ref{fig:entang} is that if the overlap between the branches is close to one, more entanglement must exist in the CM than exists in the superposition itself. This is in line with quantum mechanical intuition: the more macroscopic the superposition, the more local the CM is.

In conclusion, we have introduced the notion of a collapsing measurement (CM), which produces from an equally-weighted two-branch superposition completely distinguishable outcomes which have large overlap with one of the initial branches. When $z\ll 1$, the outcome states  are both less useful for quantum metrology and are less entangled than the superposition. We have specialized to superpositions of two product states of $N$ modes having inner product $z^{N}$ (Eq.(\ref{eqn:general})) which are macroscopic when $z\ll 1$; it is a simple task to write down a quantum superposition of non-product states that remains entangled and metrologically useful after application of the CM. For example, the state \begin{equation}
1/\sqrt{2}(\ket{\uparrow}\ket{\text{GHZ}_{N,+}} + \ket{\downarrow}\ket{\text{GHZ}_{N,-}})
\end{equation} satisfies these requirements. However, it is clear that this state is no more useful for 1-local spin Hamiltonian phase estimation than the branches it is comprised of. This fact can be used to show that such a superposition is not macroscopic \cite{dur}. In contrast, when a CM is applied to a hierarchical cat state, which has been shown to be macroscopic in Refs.\cite{volkoff,volkoff4}, the resulting product states are still useful for field displacement metrology \cite{nemoto}, despite having a quantum Cram\'{e}r-Rao bound scaling as $1/\sqrt{N}$. An example of such a state is given by:
\begin{eqnarray}\ket{\text{HCS}_{N}(\alpha)}&:=& {1\over \sqrt{2}}\left(   \left( {\ket{\alpha} + \ket{-\alpha} \over \sqrt{2+2e^{-2\vert \alpha \vert^{2}}}}  \right)^{\otimes N} \right. \nonumber \\ &+& \left. \left( {\ket{\alpha} - \ket{-\alpha} \over \sqrt{2-2e^{-2\vert \alpha \vert^{2}}}}  \right)^{\otimes N}\right) . \label{eqn:hierarchical}
\end{eqnarray}
In addition to its fundamental contribution of associating the Helstrom measurement with a physical notion of superposition collapse, the present work is expected to be useful for finding the maximal metrological usefulness of several types of multimode photonic superposition states. The experimental implementation of a CM for the state in Eq.(\ref{eqn:general}) would be a great achievement in the quantum control of macroscopic quantum systems.

\acknowledgments

I thank my Ph.D. advisor, K. Birgitta Whaley, for many stimulating conversations that helped to motivate the present research. Partial financial support was provided by the Howard H. Crandall Fellowship at UC Berkeley.

\bibliography{collapserefs}
\end{document}